\documentclass[journal, two column]{IEEEtran}
\usepackage{amsmath,amssymb}
\usepackage[dvips]{graphicx}
\usepackage{amsfonts}
\usepackage[mathscr]{eucal}
\usepackage{latexsym}
\usepackage{amsthm}
\usepackage{exscale}
\usepackage[mathscr]{eucal}
\usepackage{bm}
\usepackage[dvipsnames]{color}
\usepackage{cases}
\usepackage{color}
\usepackage{epsfig}
\usepackage{algorithm}
\usepackage{algorithmic}
\usepackage[verbose,nospace,sort]{cite}
\usepackage{tabularx}
\usepackage{multirow}
\usepackage{multicol}
\usepackage{balance}
\usepackage{caption}
\usepackage{subcaption}
\usepackage{setspace}
\usepackage{url}

\graphicspath{{./figs/}}

\begin{document}
\title{Towards Environment-Aware 6G Communications via Channel Knowledge Map}
\author{Yong~Zeng and Xiaoli Xu\\
\thanks{The authors are with the National Mobile Communications Research Laboratory, Southeast University, Nanjing 210096, China. Y. Zeng is also with Purple Mountain Laboratories, Nanjing 211111, China (e-mail: \{yong\_zeng, xiaolixu\}@seu.edu.cn).}
}
\maketitle

\begin{abstract}
This article proposes the concept of \emph{channel knowledge map} (CKM) as an enabler towards environment-aware wireless  communications. CKM is a site-specific database, tagged with the locations of the transmitters and/or receivers, that contains whatever channel-related information useful to enhance environmental-awareness  and facilitate or even obviate sophisticated real-time channel state information (CSI) acquisition. Therefore, CKM is expected to play an important role for 6G  networks targeting for super high capacity, extremely low latency, and ultra-massive connectivity, by offering potential solutions to practical challenges brought by the drastically increased channel dimensions and training overhead. In this article, the motivations of environmental-awareness enabled by CKM are firstly discussed, followed by the key techniques to build and utilize CKM.  In particular, it is highlighted that CKM is especially appealing for four channel types: channels for yet-to-reach locations, channels for non-cooperative nodes, channels with large dimensions, and channels with severe hardware/processing limitations. Two case studies with extensive numerical results are presented to demonstrate  the great potential of environment-aware communications enabled by CKM.
\end{abstract}

\section{Introduction}
Radio propagation environment critically affects the performance of wireless communication systems, and its impact is typically studied via the appropriate mathematical modelling of wireless channels--a prerequisite line of research that is of paramount importance for the design, analysis, and optimization of wireless communication networks. The most common channel modelling approach is stochastic modelling \cite{1203}, which is based on probability distributions on certain channel parameters or components, such as the shadow fading, small-scale fading, angle-of-arrival (AoA)/departure (AoD), or on those surrounding environments, like building density, heights, and sizes, etc. While stochastic channel modelling is scalable and mathematically tractable, it is usually based on the assumption that no prior knowledge about the channel environment is available, except for some high-level attributes such as the environment types, e.g., urban, suburban, rural or hilly. As such, the resulting channel model typically only depends on the relative positions of the communicating nodes, e.g., inter-node distances, while irrespective of actual environment.

\begin{figure*}
\centering
\includegraphics[width=0.9\linewidth]{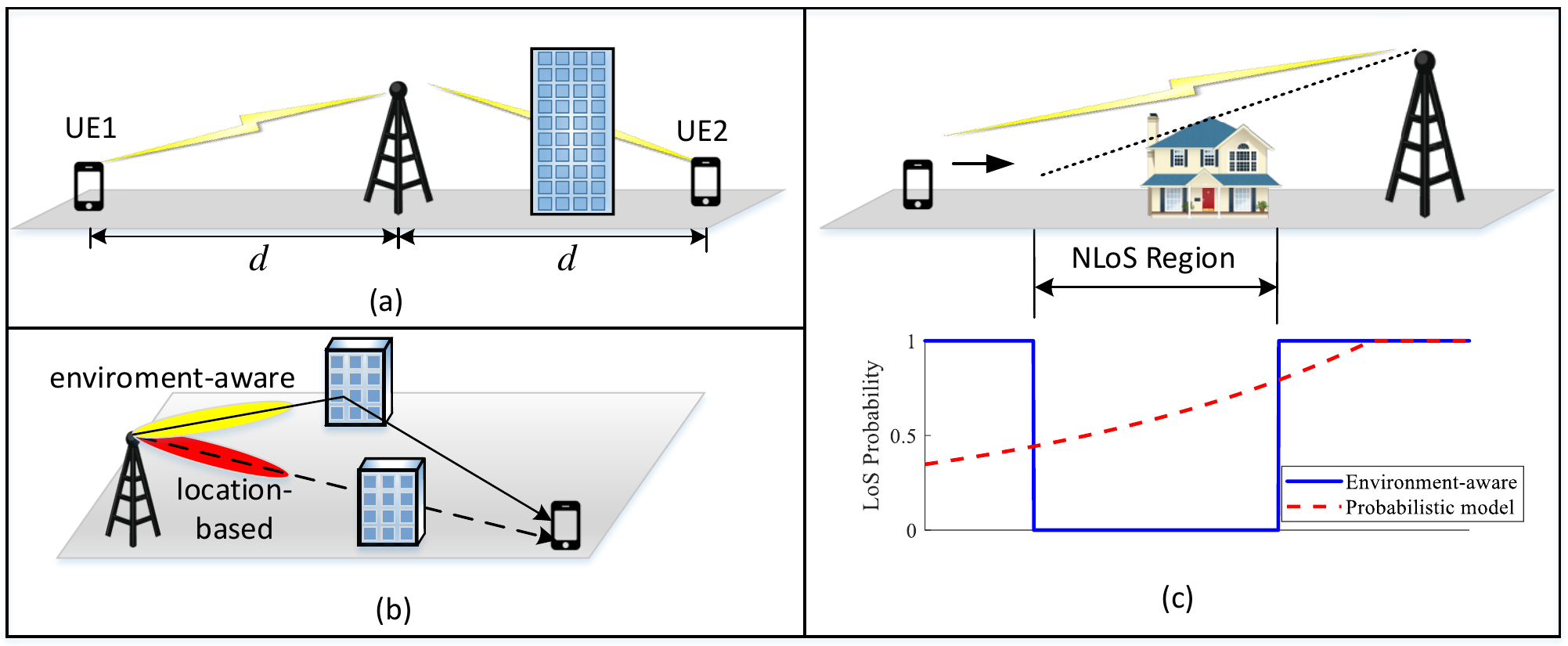}
\caption{Three examples for illustrating the importance of  environmental-awareness: (a) environment-aware channel gain prediction; (b) environment-aware versus location-based training-free beamforming; (c) environment-aware versus probabilistic LoS prediction.}\label{F:ToyExample}
\end{figure*}

To illustrate how environmental-awareness might impact wireless communication systems, we consider three simple examples. Fig.~\ref{F:ToyExample}(a) shows two user equipments (UEs) that have equal distances with the base station (BS), but a building is located between UE2 and the BS. With the classical distance-dependent path loss model, one tends to conclude that UE1 and UE2 have identical channels with the BS (at least statistically), which is obviously not true. In fact, in order to draw a more reasonable conclusion that UE2 has a worse channel than UE1, one only needs to know the BS and UE locations, together with the attributes (like the location and size) of the building. Such information is either static or readily available with built-in sensors, such as GPS receivers. In other words, with environmental-awareness, a communication system is able to more accurately predict the channels based on the location information alone, before applying any sophisticated channel acquisition technique, such as pilot-based training. As another example, Fig.~\ref{F:ToyExample}(b) shows two {\it training-free beamforming} schemes, namely {\it location-based beamforming}, which simply directs the signals towards the UE location, and the {\it environment-aware beamforming}, for which the signals are beamed towards the potential reflectors/scatterers that would eventually direct the signal to the UE. While both training-free beamforming schemes require the UE location information, the exploitation of the environment knowledge of the latter makes it possible to achieve more significant beamforming gain. In Fig.~\ref{F:ToyExample}(c), an UE is assumed to move towards the BS, and we are interested in whether there exists a line-of-sight (LoS) link along its moving path. With the prominent stochastic channel modelling, this is usually modelled as a random event, with the LoS probability modelled as a function of the relative positions of the UE and BS \cite{1214}, say the distance or elevation angle of the link, as illustrated by red dashed line in Fig.~\ref{F:ToyExample}(c). On the contrary, for the particular environment under consideration, one may easily observe that the existence/absence of LoS link is in fact deterministic, as shown by the blue solid line in Fig.\ref{F:ToyExample}(c). Such a discrepancy is mainly due to the fact that the probabilistic LoS model is meaningful only by considering a large number of realizations of the communication environments of similar type, since it does not exploit the site-specific environment knowledge. The three examples discussed above demonstrate the importance of environmental-awareness for contemporary wireless networks.

Perhaps the most straightforward approach for achieving environmental-awareness in wireless communications is to store and utilize the physical environment information, such as the three-dimensional (3D) city or terrain map \cite{1194}. However, storage of accurate environment map with fine detailed information is costly. Besides, physical  environment map alone cannot directly reflect the radio propagation characteristics, and some additional procedures like ray tracing are still needed \cite{1204}. In \cite{1203}, a map-based site-specific channel model is proposed, which is based on ray tracing using a simplified 3D geometric description of the propagation environment. However, ray tracing relies not only on the accurate geometric specification of the surrounding environment, but also their dielectric properties, which may not be always available. Besides, ray tracing algorithms usually incur high computational complexity, especially in complex urban environment, which makes them more suitable for offline numerical simulations, instead of real-time utilization for the practical design and optimization of wireless networks.

In this article, we propose the concept of {\it channel knowledge map} (CKM) as an enabler towards environment-aware wireless communications. In a broad sense, CKM is a site-specific database, tagged with the locations of the transmitters and/or receivers, that contains whatever channel-related information useful to enhance environmental-awareness and facilitate or even obviate sophisticated real-time channel state information (CSI) acquisition. One basic example of CKM is {\it channel gain map} (CGM) \cite{1206}, which is able to predict the channel gains of wireless links in a target  area. Other possible instances of CKM include channel shadowing map (CSM) that predicts the location-specific shadowing loss \cite{1062}, and channel path map (CPM) that predicts the location-specific channel path information, such as the number of significant paths and their power, phases, delays, and AoAs/AoDs, etc. Compared to the physical environment map such as 3D city or terrain map, CKM is able to directly reflect the intrinsic radio propagation environment without relying on additional parameter specification (like the dielectric properties of the environment) or complicated processing (like ray tracing). Thus, CKM  is expected to play an important role for 6G  networks targeting for super high capacity, extremely low latency, and ultra-massive connectivity \cite{1205}, by offering potential solutions to those practical challenges brought by the drastically increased channel dimensions and training overhead. Besides, it might be fair to claim that achieving environmental-awareness via techniques like CKM might be an important first step towards situation- or context-awareness for 6G.



Note that using geolocation-based database in wireless communications has received growing interest in different setups, such as spectrum database for TV white space, radio environment map for cognitive radios \cite{1215},\cite{1207}, and engineering radio map for wireless resource allocations \cite{1067}. Such existing radio maps mainly aim to provide site-specific spectrum usage information, which is thus dependent on the actual configurations and activities of the transmitters in the area. By contrast, CKM is mainly concerned about the intrinsic characteristics of the channel, which in principle only depends on the radio propagation environment, so as to achieve environmental-awareness and facilitate or even obviate real-time CSI acquisition. Besides, the following developing trends of wireless networks render CKM-enabled environment-aware communications quite promising for 6G:
\begin{itemize}
\item {\bf Convergence of localization, sensing, and communication}: CKM  primarily uses the locations of the transmitters/receivers to predict their channel knowledge, whose availability can be taken for granted for contemporary wireless systems. In fact, it is envisioned that 6G should offer an integration of localization, sensing, and communication, with centimeter-level localization accuracy \cite{1208}, i.e., around the signal wavelength scale. This implies that it is in principle possible to construct accurate CKM to predict not just the large-scale channel knowledge such as path loss and shadowing, but also the exact CSI or channel impulse response that de-correlates over wavelength scale.
\item {\bf Increased channel  dimension and imperative for cost-effective implementation}: 6G networks are expected to involve wireless channels with large dimensions, in terms of the increased number and density of the connected devices, the ever-increasing antenna array size, as well as the wider bandwidth usage. This makes it incur prohibitively high overhead if real-time CSI is acquired purely relying on pilot-based channel training and feedback techniques.  The situation gets even worse with the imperative need of cost-effective implementations like using only very simple hardware and signal processing techniques. In fact, study has revealed that for millimeter wave (mmWave) analog beamforming  with the BS having 64  antennas and UE having 16 antennas, it may take up to 5.2 seconds to scan all transmit and receive beam directions, which makes them infeasible  for practical usage \cite{1209}. Thus, developing environment-aware communication techniques like CKM to facilitate or even obviate real-time channel training is quite appealing for 6G.
\item {\bf Enhanced data acquisition, storage and mining capability}: The accuracy and practical usefulness of CKM depends on the availability and quality of the data. The abundant data generated by contemporary wireless networks, together with the recent advancement of data storage and mining capabilities, like deep learning, paves the way to construct highly efficient and effective CKM that would not be  possible before.
\end{itemize}

In this article, we provide an overview on CKM-enabled environment-aware wireless communications. The various aspects of building CKMs are discussed, including their classifications, data acquisition and learning techniques, followed by the utilization of CKM for realizing {\it training-free communications} or {\it CKM-assisted channel training}. Four channel types for which CKM is particularly useful are highlighted. Furthermore, two use cases with extensive numerical results are presented to demonstrate the great potential of environment-aware communications enabled by CKM.


\section{Building Channel Knowledge Map}
The essential task of building a CKM is to establish an accurate mapping between the channel knowledge of interest and all potential transmitter/receiver locations in the target area/space, based on a finite number of collected data tagged by locations and also possibly time. This is feasible due to the strong spatial correlations of wireless links, together with the (quasi)-static nature of many aspects that impact wireless channels, such as the deterministic blockages, reflectors, and scatters. Since CKM is site-specific, each BS or a group of adjacent BSs may construct and continuously update its own local CKM for its surrounding 2D area or 3D space. Besides, to enable CKM-based inter-cell coordination, different BSs may exchange their local CKMs via the backhaul links, say during off-peak hours.

\subsection{CKM Classification}
CKM can be classified based on different criteria. According to the types of wireless links, we may have {\it BS-to-any} (B2X) and {\it any-to-any} (X2X) CKM.  B2X CKM aims to portray the site-specific channel knowledge between a BS and its surrounding 2D or 3D location space. This is the right CKM type for the conventional BS-centric communications. Since the location of a ground BS is typically fixed once deployed, B2X CKM only requires the 2D or 3D coordinate of the potential UE locations as the input. Note that the map area/space of B2X CKMs associated with different BSs may overlap, which is necessary to facilitate CKM-based cell association, handover, and inter-cell interference coordination (ICIC). On the other hand, X2X CKM aims to offer site-specific channel knowledge between any pair of transmitter-receiver locations for the area/space of interest. Therefore, X2X CKM is at least 4D or 6D. X2X CKM is important for contemporary wireless networks with device-to-device (D2D) communications, distributed antenna systems, and cell-free networking architectures, where both the transmitters and receivers might be at arbitrary locations. Besides, even for the conventional BS-centric communication, X2X CKM is also quite useful for advanced transmission technologies, such as in-band uplink-downlink full-duplexing, where an UE in uplink communication may utilize the CKM to predict its interference  caused to a co-channel UE in downlink. Besides, due to the heterogenous nature of contemporary wireless networks, it is likely that both B2X and X2X CKMs need to be maintained to complement each other.

On the other hand, based on the types of channel knowledge to be learned, CKM may be classified as {\it quasi-stationary} or {\it volatile}. Quasi-stationary CKM aims to predict the channel knowledge that remains relatively stable, or only vary at large time scales. Typical examples include the path loss, shadowing, presence/absence of LoS links, number of deterministic paths and their powers/phases/delays/AoAs/AoDs. Quasi-stationary CKM may also include the large-scale location-specific (instead of site-specific only) channel statistical parameters like the angular and delay spread, Rician factors, etc. Quasi-stationary CKM requires relatively coarse localization accuracy, say comparable to the channel shadowing coherent distance. One might argue that as quasi-stationary channel knowledge only varies slowly, it only requires infrequent training, and thus incurs little overhead. This seems to render it unnecessary to build quasi-stationary CKM. While such arguments might be true for the simple point-to-point communication links, for future wireless networks involving large channel dimensions and diversified links, we may no longer take the large-scale channel knowledge for granted, as evident by the four typical channel types discussed in Section~\ref{sec:ChannelTypes}. On the other hand, volatile CKM involves channel knowledge that varies  more frequently, usually due to the varying of the propagation environment that is difficult to predict beforehand. Typical examples include the instantaneous CSI due to multi-path fading and Doppler  frequency shift, the channel angular/delay profile or the channel impulse response. While it is challenging to construct volatile CKM for high mobile environment, with the continuous improvement of localization accuracy and advanced data acquisition and processing capabilities, it is becoming more feasible to develop volatile CKM for relatively static or periodically changing environment, like automatic factory in Industrial Internet of Things (IIoT) applications.

\subsection{Data Acquisition and CKM Learning}

The data used to build a CKM can be acquired via offline numerical simulation and offline or online measurements. With offline simulation, location-specific channel knowledge could be generated based on ray tracing with the available physical environment information. 
 While offline simulation-based data acquisition is cost-effective for implementation, the resulting data quality heavily depends on the accuracy and detailedness of the knowledge on the physical environment, which limits  their accuracy. This thus calls for on-site channel measurements for data collection. On one hand, dedicated measurement devices such as ground or aerial vehicles  could be dispatched offline for data collection exclusively. On the other hand, data could also be collected online concurrently when actual communication takes place. With contemporary wireless networks that involve extensive channel training, but with typically one-shot usage of the trained result, the system only needs to introduce one additional step to store the location-tagged channel training result for online CKM data collection. 
Compared to offline numerical-based simulation, the offline/online measurement-based data collection is expected to provide more  accurate data  that better reflects the reality of radio propagation environment, though with higher implementation cost. Thus, in practice, a combination of the three aforementioned  data collection methods could be used to complement each other, e.g., building a coarse CKM based on offline numerical simulation and refining it with offline/online measurements. For all the data acquisition methods, a fundamental question is how dense and how frequent the data should be sampled along the space and time dimensions. Intuitively, this depends on the channel knowledge type of interest and its spatial/temporal coherence distance/time. For example, for large-scale channel knowledge such as CSM, data sampling on meter-level is typically sufficient, while that for volatile  CKM is usually much shorter.

In terms of the mathematical representation of CKM, the simplest approach is table-based, i.e., all collected data is stored in a table, whose query input is the 2D/3D UE location for area/space B2X CKM and 4D/6D transmitter-receiver location pairs for area/space X2X CKM. On the other hand, for volatile CKM or CKM where wideband frequency-dependent knowledge if of interest, the time and frequency should be also added as query input dimensions. Thus, we may have CKMs of up to 8 input dimensions.  To reconstruct the entire CKM based on the finite number of collected data, the  interpolation techniques such as Kridging-based method could be applied \cite{1210}. However, table-based CKM requires high storage capacity and query response time due to the ``curse of dimensionality" issue. For example, for an 8D CKM with even very coarse 10-level quantization on each dimension, we would already have  $10^8$ entries to store and query the table. Besides, table-based CKM does not really try to learn the intrinsic structure or feature of the data, but instead simply ``memorize" it. Therefore, an alternative approach is to learn the CKM based on deep neural networks (DNNs), with the transmitter/receiver location, time, and frequency as the DNN input, and the channel knowledge to be predicted as the output. As a result, the powerful deep learning algorithms such as backpropagation can be applied to train the coefficients of the DNN with the collected data. Compared to table-based CKM, DNN-based CKM resolves the ``curse of dimensionality" problem and is more responsive for channel knowledge prediction once properly trained. Note that there are different variations for the DNN-based CKM architecture. For example, for CKM with frequency-selective channel knowledge, if we are only interested in the channel knowledge of a few sub-bands, then we may move the frequency dimension from the DNN input to output by expanding the output dimension accordingly. One such example is discussed in Section~\ref{sec:D2D}.

\section{Utilizing Channel Knowledge Map towards Environment-Aware Communications}
CKM can be utilized for many purposes whenever channel knowledge is helpful, ranging from network planning such as cell site selection, on-demand relay positioning, coverage enhancement and cell association, to adaptive communications like dynamic resource allocation, interference mitigation, channel assignment, beam selection/aligment, user pairing for D2D and non-orthogonal multiple access (NOMA), and proactive mobility management, etc. In particular, CKM is especially appealing for the following four channel types.

\subsection{Typical Usage Scenarios}\label{sec:ChannelTypes}
{\bf Channels for yet-to-reach locations}:  Existing methods for CSI acquisition mainly rely on pilot-based channel training,  which, however, is possible only when the communication devices actually reach the locations for the channels to be trained. In contemporary wireless networks, it is often useful to acquire the channel knowledge for those wireless links with {\it yet-to-reach} or {\it never-to-reach} locations to gain a global channel view, so as to enable predictive network planning or resource allocation. This is possible by invoking the channel prediction capability of CKM. For example, by utilizing CKM together with UE trajectory prediction, foresighted handover mechanisms could be developed to resolve issues like ``ping-pong'' handover in complex environment. As another example, with CKM over the 3D  aerial space, a cellular-connected UAV  may plan its flying trajectory to avoid coverage holes in the sky \cite{1212}. Such tasks are difficult, if not impossible, to accomplish if purely relying on conventional training-based channel estimation.

{\bf Channels for non-cooperative nodes}: Typical examples for channels with non-cooperative nodes include: (i) the wireless links between secondary transmitters and primary receivers in cognitive radio systems; and (ii) the wireless links between legitimate transmitters and adversary eavesdroppers for physical layer security communications. The non-cooperative nature of such nodes makes it difficult to acquire their CSI with conventional channel training approach, though they are essential for interference mitigation and security protection. CKM offers a promising solution to resolve this issue, by predicting the channel knowledge mainly based on their location information, which is more readily available, say with passive sensing for eavesdroppers or directly retrieved from the network database for primary receivers.

 {\bf Channels with large dimensions}: The overhead of channel training and feedback, in terms of communication resources like training energy, pilot sequence length, and response delay,  increase drastically with channel dimensions. 6G is expected to involve communications with extremely large antenna arrays, super high frequency (up to Tera-Hertz), wide bandwidth, and ultra-massive connectivity, but at the same time targeting for sub-milliseconds (ms) communication latency. This brings a big challenge for conventional pilot-based channel training. On the other hand, CKM offers a promising solution to facilitate or even obviate real-time CSI acquisition, which is quite appealing for simultaneously achieving high-capacity and low-latency communications. 

 {\bf Channels with severe hardware/processing limitations}: The large channel dimension together with the imperative needs for energy- and cost-effective implementation of 6G requires using simple hardware and low-complexity signal processing techniques, such as analog or hybrid analog/digital beamforming  with only one or very few radio frequency (RF) chains, low-resolution analog-to-digital converters (ADCs), or even fully passive transmission/reflection with reconfigurable intelligent surfaces \cite{1213}. Such restrictive transceiver architectures exacerbate the training and feedback overhead of conventional pilot-based training for channels  with large dimensions. In this case, CKM-assisted wireless network is expected to provide a promising solution by exploiting the environmental-awareness for reducing or even avoiding training overhead.

\subsection{Training-Free Communications and  CKM-Assisted Channel Training}
 Depending on the specific applications and performance requirements, CKM can be utilized in two different ways: {\it training-free communications} and {\it CKM-assisted channel training}.

 With CKM-enabled training-free communications, the channel knowledge predicted by the CKM is directly used for the system design and performance optimization, without requiring  the conventional pilot-based channel training. This is usually applicable when the CKM is sufficiently accurate and/or when only coarse performance guarantee is needed. Some possible examples include power allocation, interference mitigation, cell  association, and user pairing based on large-scale channel gains (e.g., path loss and shadowing).  In certain scenarios, the CKM-predicted channel knowledge is firstly used to reconstruct the channel before being used, an example of which is discussed in Section~\ref{sec:mmWavBF}. CKM-enabled training-free communication is especially appealing for low-complexity and low-latency communications since it eliminates the overhead and processing delay associated with channel training and feedback.

\begin{figure}
\centering
\includegraphics[width=0.95\linewidth]{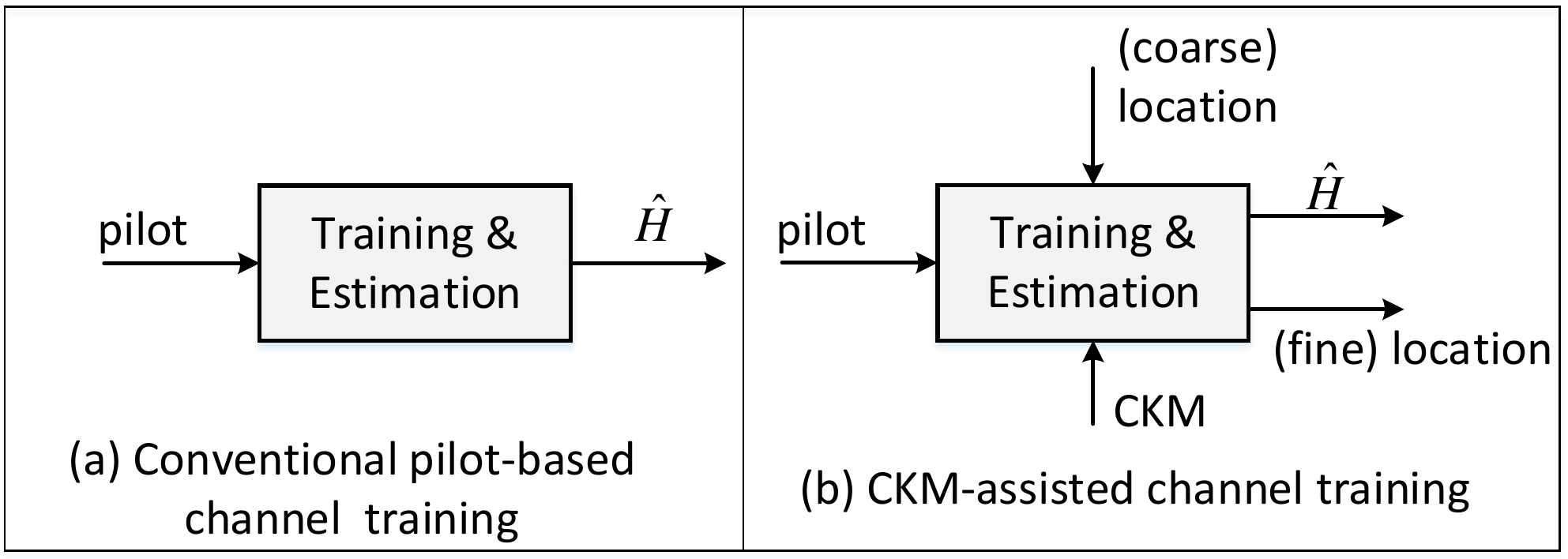}
\caption{Conventional pilot-based channel training versus CKM-assisted training with CKM and node location as additional side information.}\label{F:ChannelTraining}
\end{figure}

 On the other hand, with CKM-assisted channel training, CKM is used as additional side information to enhance the performance of conventional pilot-based channel training, as illustrated in Fig~\ref{F:ChannelTraining}. This is usually the case  when only coarse CKM is available and/or when strict performance guarantee is needed. Besides CKM, another side information that is useful for channel training is the (possibly coarse) location information of the transmitter and/or receiver, which is readily available with variety of localization techniques today, such as GPS, vision/light-based localization. Note that not only the communication channel, but also refined node locations can be estimated with training. Since side information at least does not hurt, it is expected that CKM-assisted channel training with environmental-awareness in Fig.\ref{F:ChannelTraining}(b) is superior than the conventional pilot-based  channel training in Fig.~\ref{F:ChannelTraining}(a), in terms  of  the performance  enhancement and overhead reduction. For instance,  for the mmWave beamforming example shown in Fig.~\ref{F:ToyExample}(b), the conventional environmental-ignorant  beam training would require exhaustive beam sweeping over the entire 3D space. By contrast, the searching space can be significantly reduced with CKM-assisted beam training, thanks to its environmental-awareness.

\begin{figure*}
\centering
\begin{subfigure}{0.48\linewidth}
\centering
\includegraphics[width=0.6\linewidth]{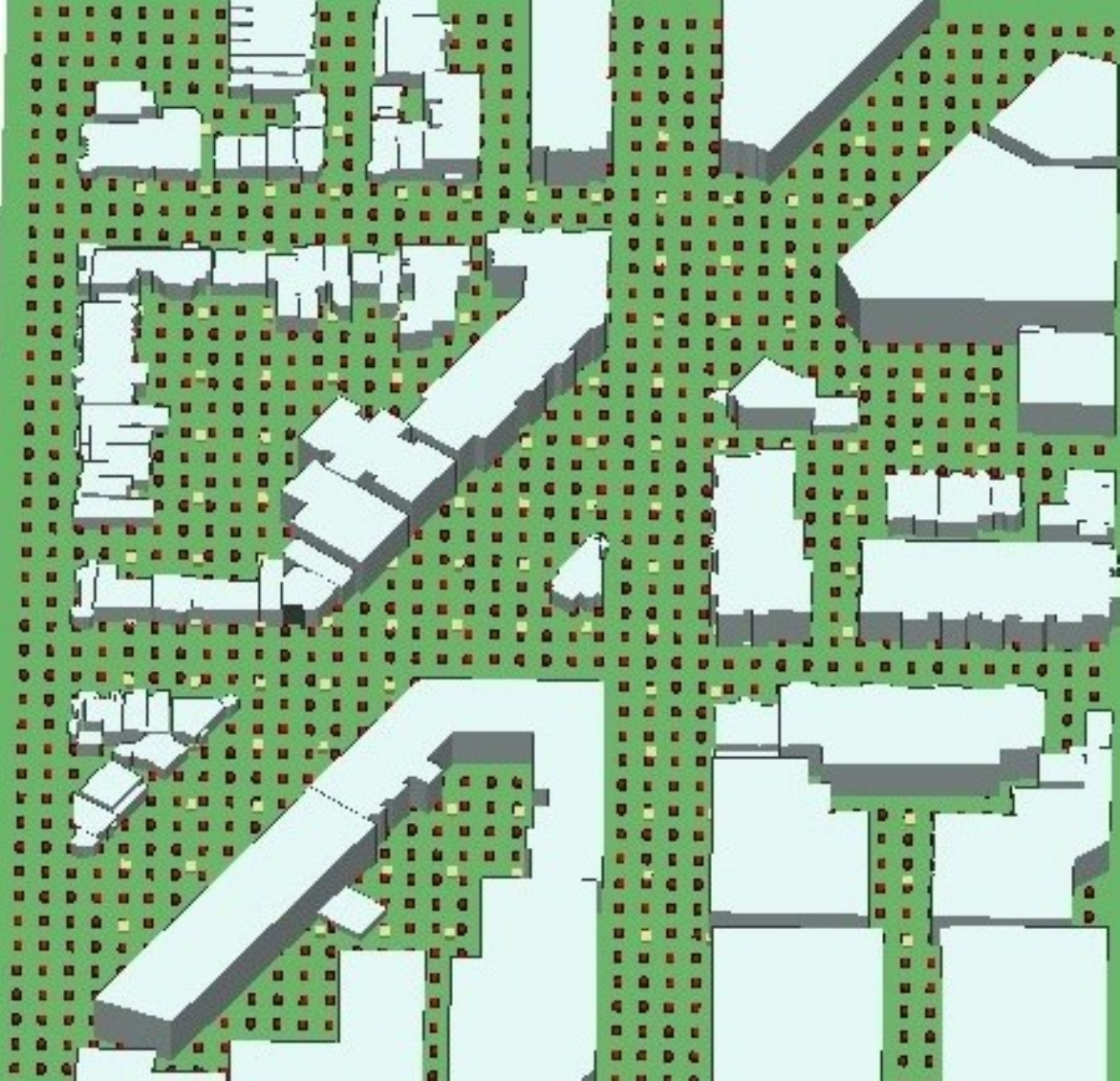}
\caption{3D physical environment. Small squares denote example sampled transmitter/receiver locations for ray tracing data collection.}
\end{subfigure}\hspace{6pt}
\begin{subfigure}{0.48\linewidth}
\includegraphics[width=0.95\linewidth]{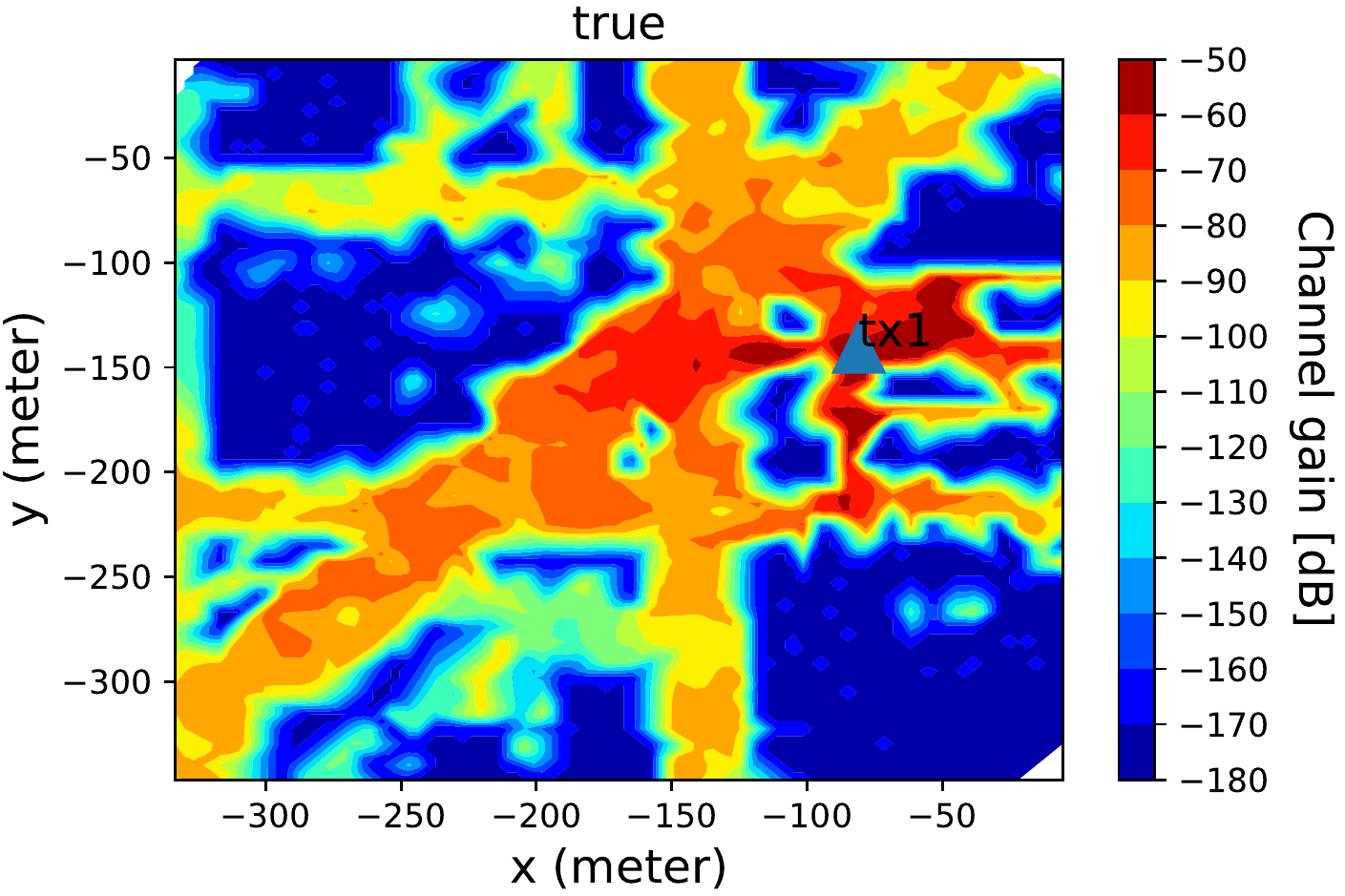}
\caption{True}
\end{subfigure}
\\
\begin{subfigure}{0.48\linewidth}
\includegraphics[width=0.95\linewidth]{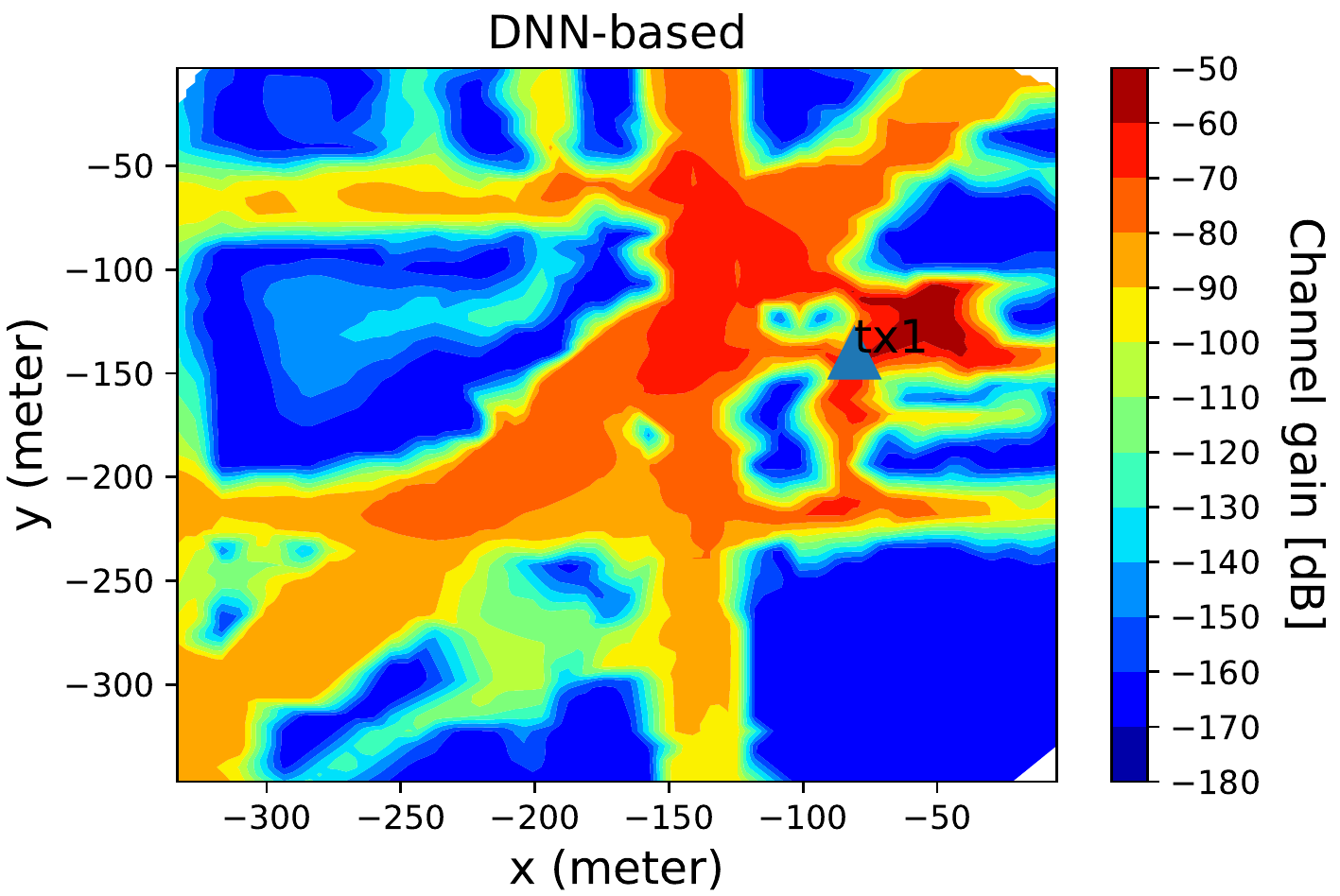}
\caption{DNN-based}
\end{subfigure}\hspace{2pt}
\begin{subfigure}{0.48\linewidth}
\includegraphics[width=0.95\linewidth]{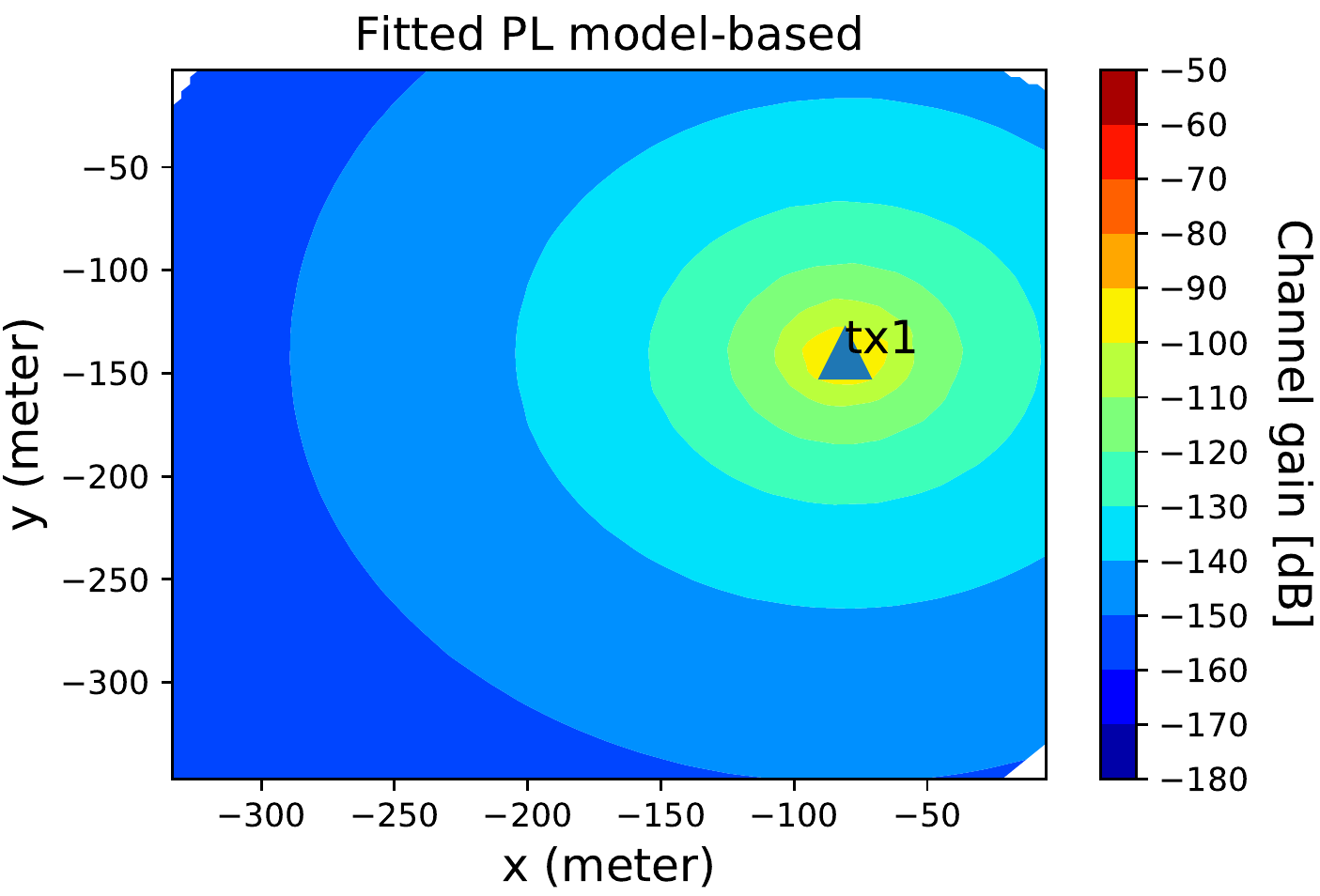}
\caption{Fitted path loss model-based}
\end{subfigure}
\caption{3D physical environment and CGMs for D2D sub-band assignment of one example transmitter location and sub-band.}\label{F:D2D}
\end{figure*}

\section{Case Studies and Numerical Results}
To evaluate the performance of environment-aware communications enabled by CKM,  two concrete examples are studied: CKM-based D2D sub-band assignment and CKM-based mmWave  beam selection. The simulation source codes and  all relevant data are available online at Github\footnote{\url{https://github.com/xuxiaoli-seu/Environment_Aware_Communications}}.

\subsection{CKM-based D2D Sub-band Assignment}\label{sec:D2D}
D2D communication enables devices that are in close proximity to exchange information directly, without having to traversing the BS or core network. Consider an area consisting of $K$ D2D user pairs that need to share $N$ frequency sub-bands, where $N<K$. One fundamental problem is sub-band assignment, i.e.,  choosing one out of the $N$ sub-bands for each D2D pair, such that the achievable sum rate of all D2D users is maximized. Clearly, the solution depends on the available channel knowledge, not just of those direct links between a D2D transmitter and its desired receiver, but also those cross links, since D2D users assigned to the same sub-band would  interfere  with each other. Therefore, with pilot-based channel training, a total of $K^2N$ channels need to be estimated, which would incur high training overhead  and complex  handshaking processes when $K$ and/or $N$ is large. This issue could be circumvented by CKM-enabled environment-aware D2D sub-band assignment.

The 3D physical environment of the studied urban area for D2D communications is shown in Fig.~\ref{F:D2D}(a). A X2X CGM is constructed to predict the channel gains for each of the $N$ sub-bands, with the transmitter-receiver location pair given as the input.  The CGM is trained by a fully connected feedforward DNN with one input layer of dimension $4$, corresponding to the transmitter-receiver pair locations, one output layer of dimension $N$, corresponding to the predicted channel gains of the $N$ sub-bands, and 7 hidden layers that have 64, 128, 256, 512, 256, 128, 64 neurons, respectively. The standard ReLU activation is used for each hidden layer, and the DNN is trained to minimize the mean squared error (MSE) between the predicted and true channel gains. The commercial ray tracing software Remcom  Wireless Insite\footnote{https://www.remcom.com/wireless-insite-em-propagation-software} is used to simulate the actual radio propagations and generate the training and test data. Fig.~\ref{F:D2D}(b)-(d) shows the resulting DNN-based CGM associated with one example transmitter location and sub-band. As a  benchmark  comparison, we also consider the resulting CGM with the classical path loss model based channel gains, where the modelling parameters  including the path loss exponent, intercept, and frequency-dependent factor are curve-fitted based on the same training data as the DNN-based approach. It is observed from Fig.~\ref{F:D2D} that the DNN-based CGM matches quite well with the true map, and they both follow the layout of the physical environment in Fig.~\ref{F:D2D}(a). By contrast, the fitted path loss model-based CGM results in concentric contours for channel gains, which is far from the reality, due to the restriction of the theoretical  path loss model that only depends on node distances, rather than their specific locations.

\begin{figure}
\centering
\includegraphics[width=0.95\linewidth]{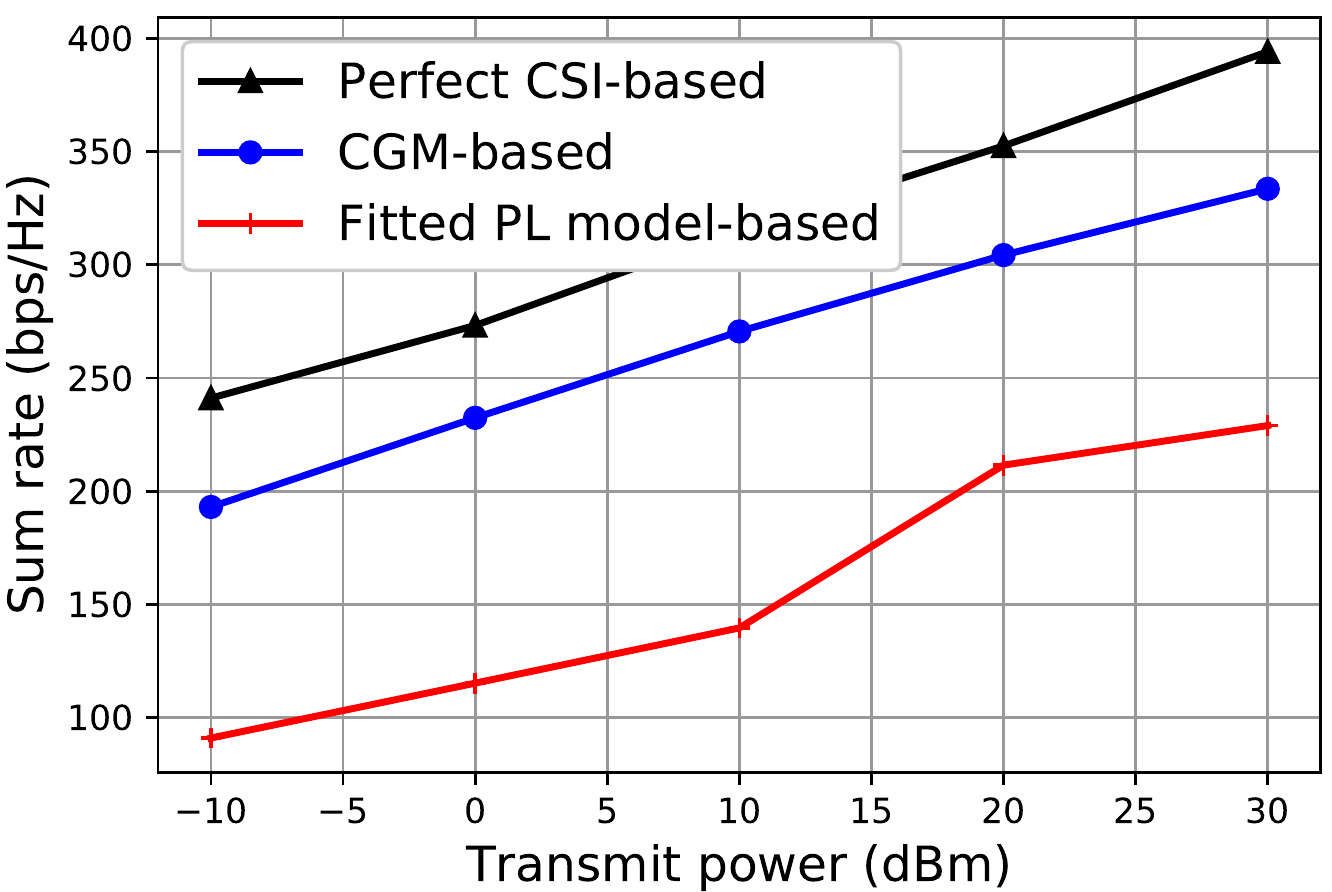}
\caption{Communication sum rate with CKM-based D2D  sub-band assignment versus benchmark schemes..}\label{F:D2Dsumrate}
\end{figure}

Fig.~\ref{F:D2Dsumrate} shows the achievable sum rate with D2D sub-band assignment based on perfect CSI, CGM, and fitted path loss model, respectively, where $K=30$ and $N=12$. Note that since the sub-band assignment is a combinatorial optimization problem, which  is NP hard in general, we apply a greedy-based assignment algorithm to all the three schemes, i.e., the $K$ D2D users are considered successively, and for each D2D user under consideration, the sub-band that would lead to the highest sum rate is selected. It is observed that the CGM-based assignment significantly outperforms that based on fitted path loss model. This is expected since the former is able to predict the channel gains more accurately, as evident from  Fig.~\ref{F:D2D}. The figure also shows that the CGM-based scheme performs closely to that based on perfect CSI, yet without requiring any channel training. This demonstrates the great potential of CKM-based resource allocation for wireless networks.

\subsection{CKM-based mmWave Beam Selection}\label{sec:mmWavBF}
\begin{figure*}
\centering
\begin{subfigure}{0.45\linewidth}
\centering
\includegraphics[width=0.7\linewidth]{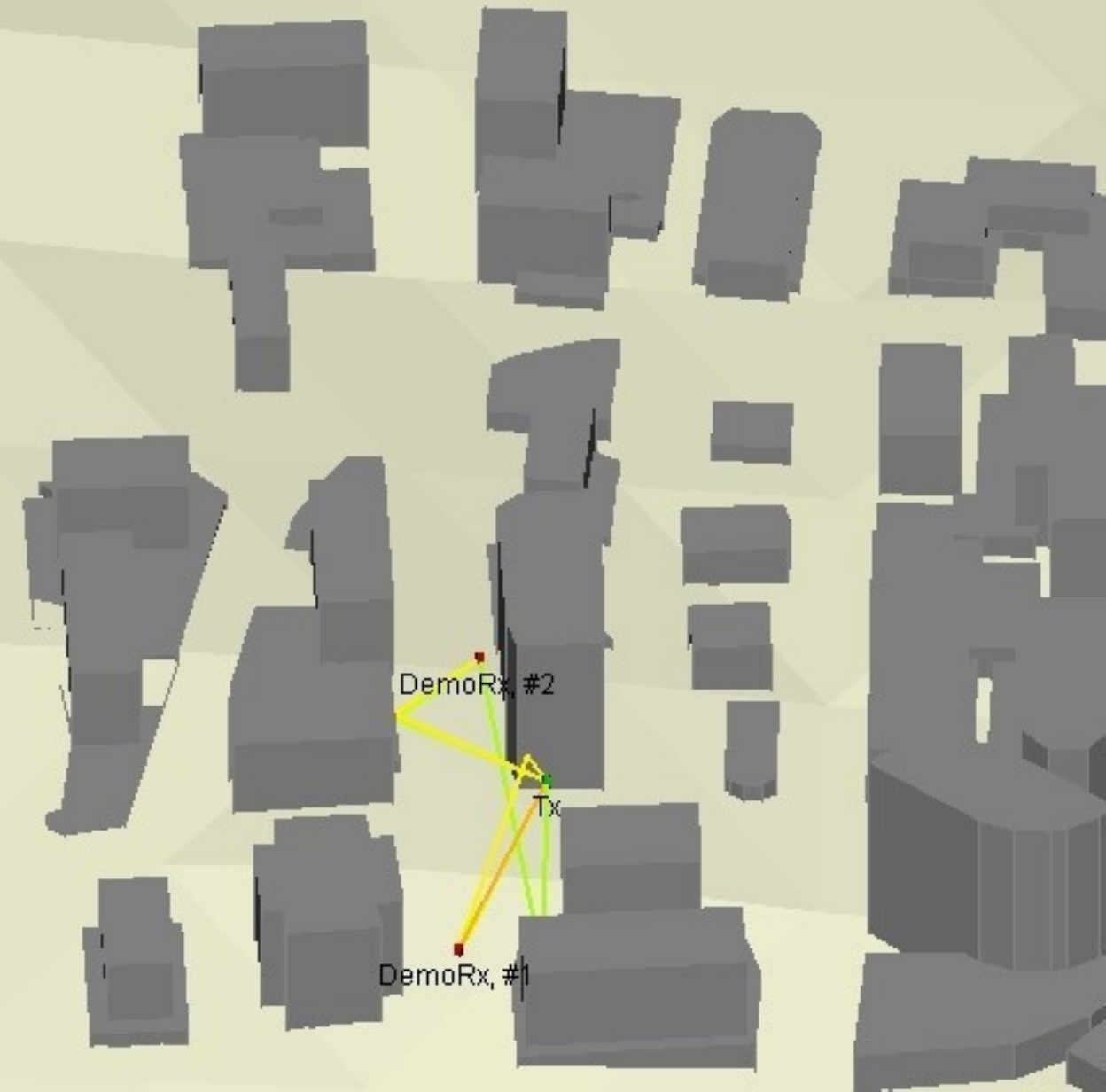}
\caption{3D physical environment. The BS transmitter and two example receivers are labelled, together with their three strongest paths.}
\end{subfigure}\hspace{2pt}
\begin{subfigure}{0.45\linewidth}
\includegraphics[width=0.95\linewidth]{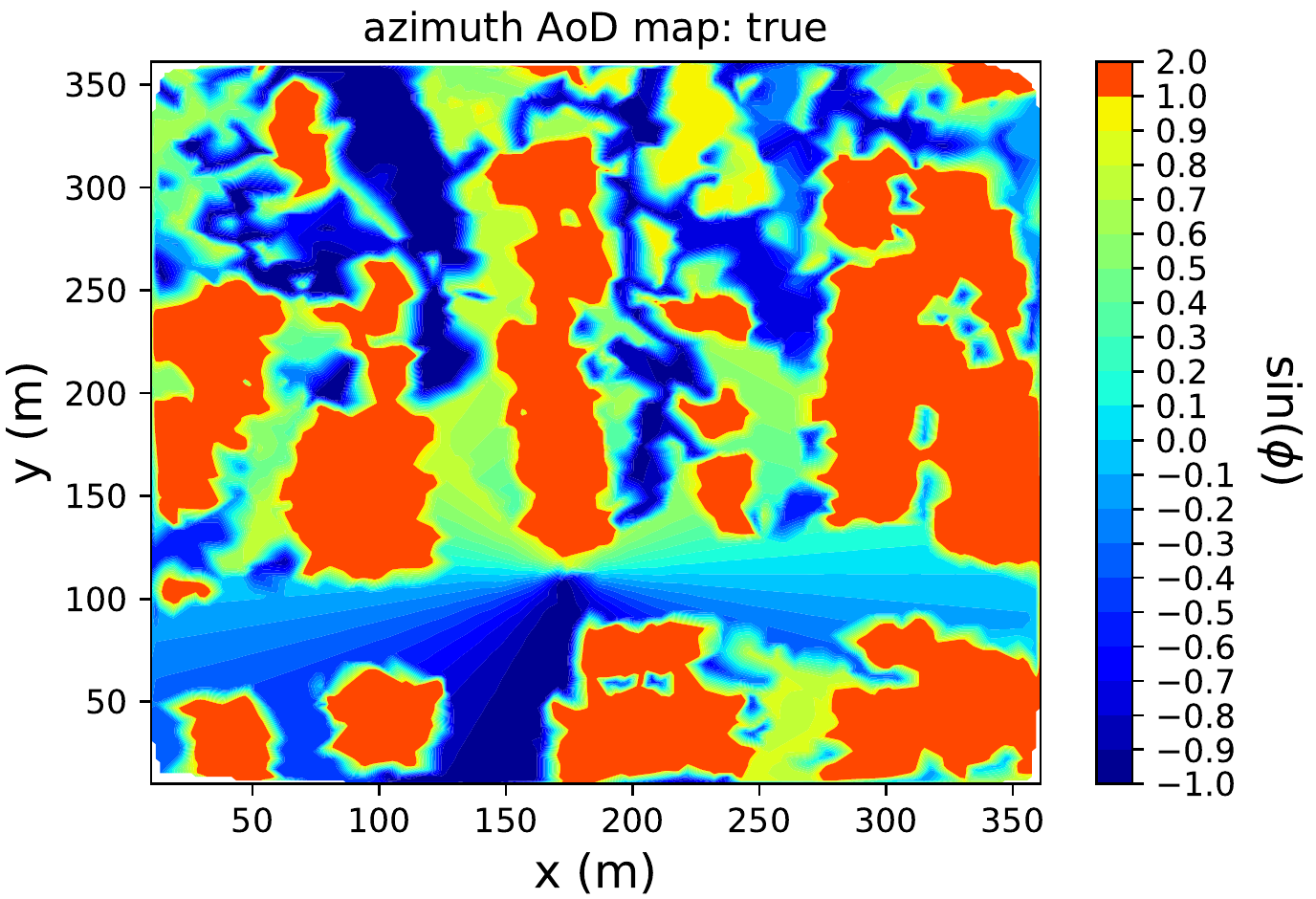}
\caption{True}
\end{subfigure}
\\
\begin{subfigure}{0.45\linewidth}
\includegraphics[width=0.95\linewidth]{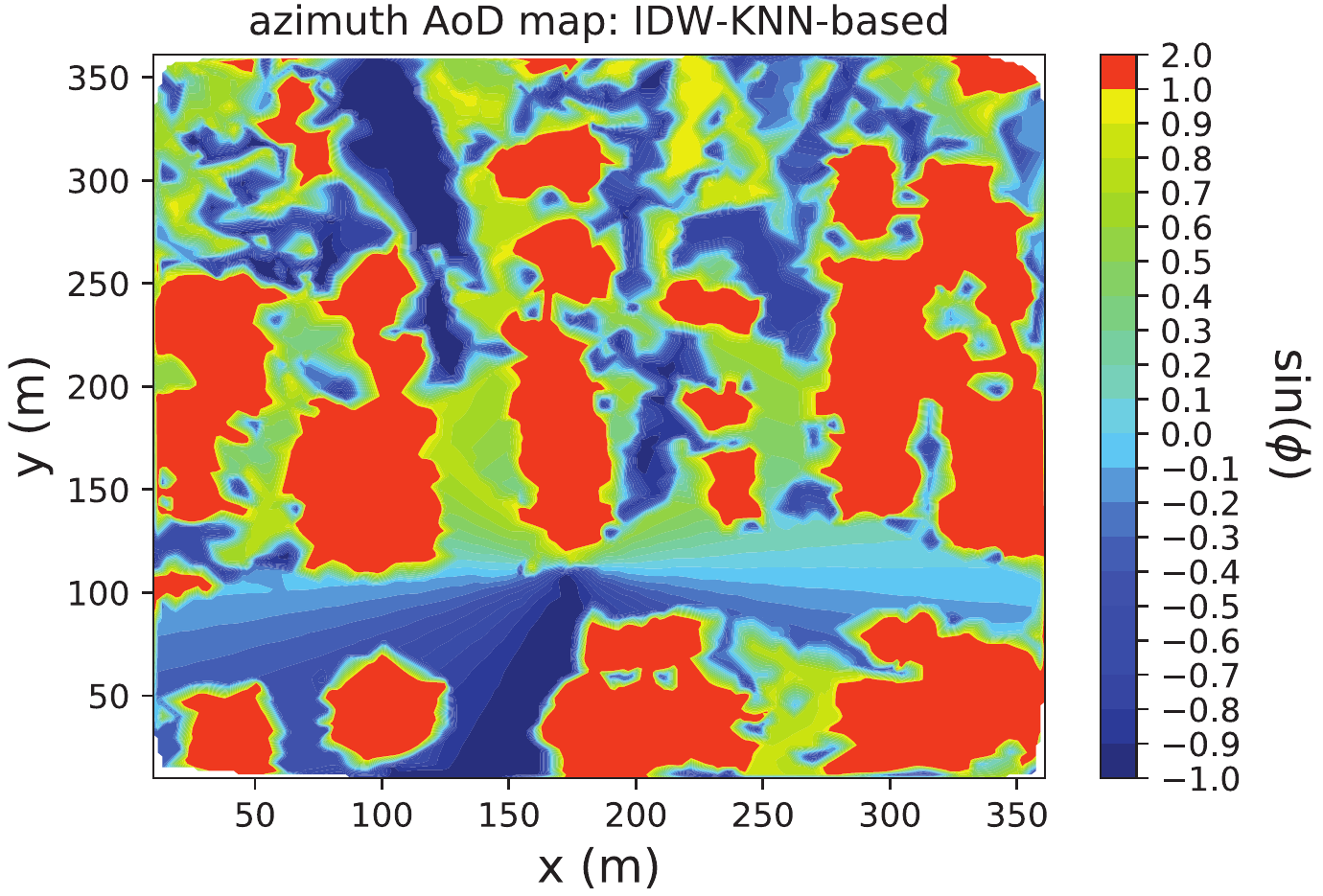}
\caption{IDW KNN-based}
\end{subfigure}\hspace{2pt}
\begin{subfigure}{0.45\linewidth}
\includegraphics[width=0.95\linewidth]{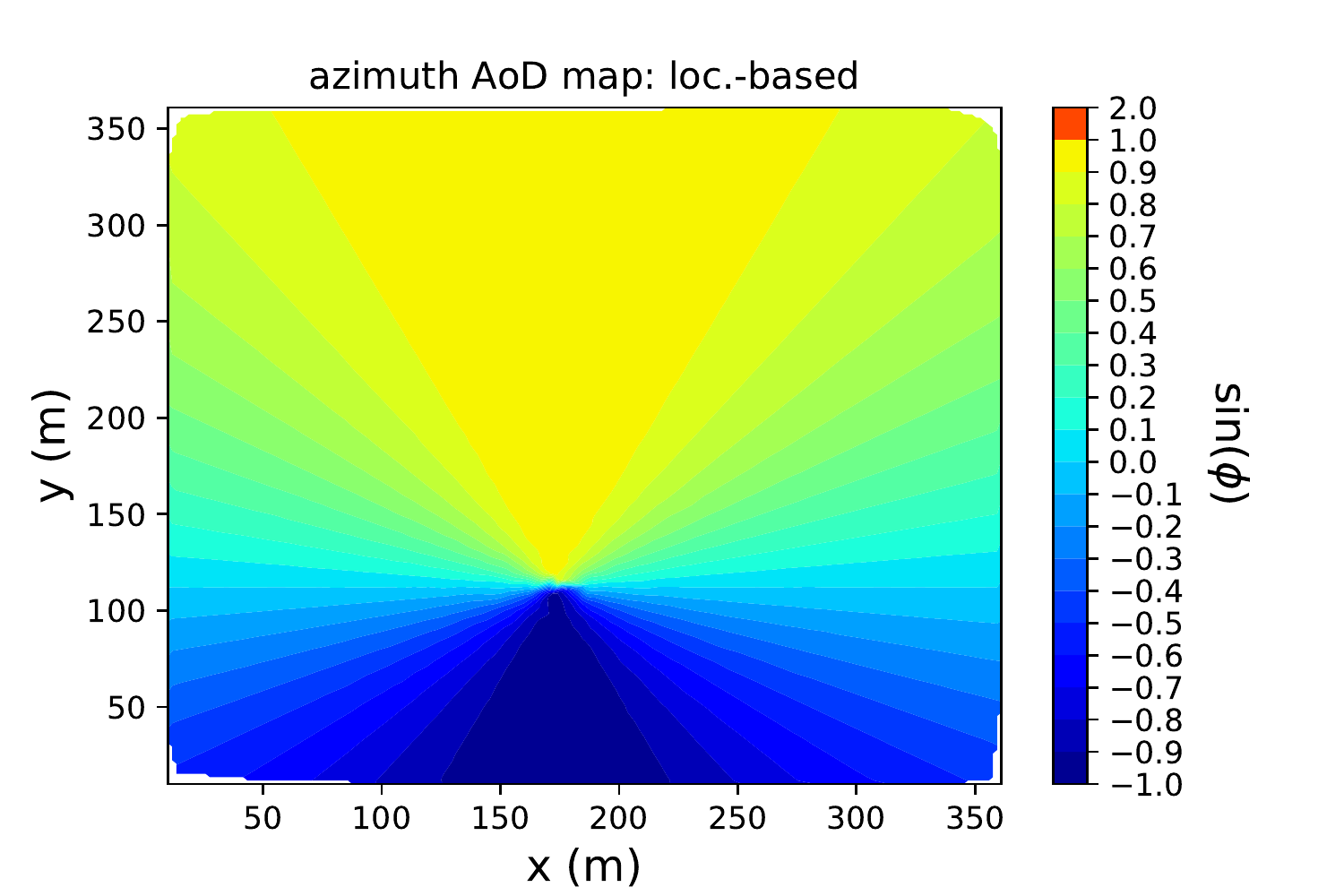}
\caption{Location-based}
\end{subfigure}
\caption{3D physical environment and azimuth AoD maps of the strongest path for mmWave  beam selection. $\sin(\phi)=2$ signifies areas without any detectable channel path.}\label{F:mmWave}
\end{figure*}

For the second example, we consider a mmWave communication system in an urban environment shown in Fig.~\ref{F:mmWave}(a). To enable training-free mmWave beamforming, a B2X CPM is constructed to predict the essential information of the significant channel paths. In our numerical results, the output of the CPM include the path gain, phase, zenith AoD $\theta$ and azimuth AoD $\phi$ of the three strongest paths. Thus, the input and output dimensions of the CPM is 2 and 12, respectively. The Remcom commercial ray tracing software Wireless Insite is used to generate the groundtruth data, which is stored in a table since the query input dimension is small. Based on the finite ray tracing data samples, the inverse distance weighting (IDW) method of the $K$ nearest neighbors (KNN) is used to construct the entire CPM of all locations. With $K=3$ nearest neighbors used, Fig.~\ref{F:mmWave} shows the resulting azimuth AoD maps of the strongest path. As a comparison, the location-based  AoD map is also shown, i.e., the AoDs are simply calculated based on the relative positions of the BS  and receiver,  while irrespective of the propagation environment. It is observed that the KNN-based CPM matches quite well with the ground-truth map. By contrast, the location-based azimuth AoD map is given by the simple rays, which is far from  reality due to its ignorance of the actual radio propagation environment.

\begin{figure}
\centering
\includegraphics[width=0.99\linewidth]{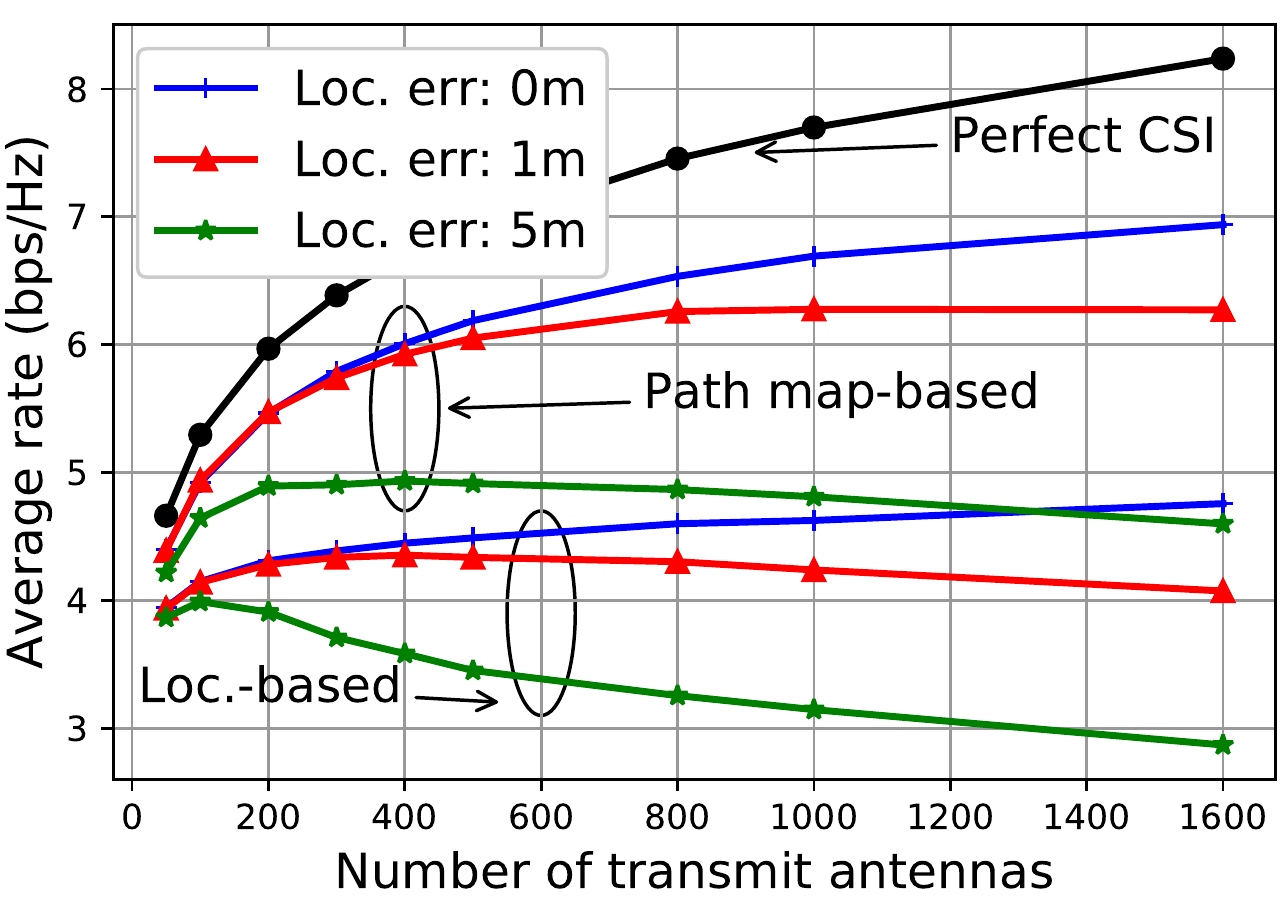}
\caption{Communication rate with CKM-based mmWave beam selection versus benchmark schemes.}\label{F:mmWaveAvgRate}
\end{figure}

Fig.~\ref{F:mmWaveAvgRate} shows the average communication rate versus the number of BS transmit antennas with mmWave beam selection based on perfect CSI, CPM, and UE locations, respectively. The average is taken with respect to $1000$ randomly selected UE locations that have at least one detectable channel path. The BS is assumed to be equipped with $N_z \times N_y$ uniform planar array (UPA) with adjacent elements separated by half wavelength,  where $N_z$ and $N_y$ represent  the number of vertical and horizontal elements, respectively. We assume that $N_z$ is fixed to $10$ while $N_y$ varies. The low-complexity codebook-based analog beamforming is performed, where the codebook is generated based on the transmit array response vectors with uniformly sampled values over $\sin(\phi)$  and  $\cos(\theta)$, with resolution $1/(2N_y)$ and $1/(2N_z)$,  respectively. For CPM-based scheme, the channel is first reconstructed based on the predicted path information, according to which the best beam is selected. Besides the ideal case with errorless UE  location information, we also consider the practical cases with mean UE localization errors of 1 meter (m) and 5 m.  It is firstly observed that the CPM-based beam selection, which is environment-aware, significantly outperforms the location-based scheme. It is also observed that even with a practical localization error of 1 m,  the CPM-based scheme, which is training-free, can already achieve a significant portion of the performance of that with perfect CSI, e.g., around $90\%$ when the number of transmit antennas is $400$. Fig.~\ref{F:mmWaveAvgRate} also reveals that for both CPM-based and location-based beam selection, as the number of transmit antennas increases, the beamforming performance becomes more sensitive to the UE localization accuracy, which is expected due to the sharper beam formed with larger antenna arrays.

\section{Conclusion and Future Work}
In this article, we provided an overview of environment-aware wireless communications enabled by the concept of CKM, which is a site-specific database, tagged with the locations of the transmitters and/or receivers, being able to facilitate or even obviate sophisticated real-time CSI acquisitions. The motivations of CKM-enabled  environment-aware communications were discussed, together with the main techniques for building and utilizing CKM. Numerical results for CKM-based D2D sub-band assignment and mmWave beam selection were presented to corroborate our discussions. Some promising directions for further research include developing the general framework of CKM-enabled environment-aware communications, the fundamental theoretical performance analysis, the efficient construction and utilization of high-dimensional and volatile CKMs, and the study of simultaneous communication, sensing and channel knowledge mapping.

\bibliographystyle{IEEEtran}
\bibliography{IEEEabrv,IEEEfull}
%
%

\end{document}